\documentstyle{mn}
\input{epsf}

\title[Cluster mass functions in the quintessential Universe]
      {Cluster mass functions in the quintessential Universe}

\author[E. L. {\L}okas, P. Bode and Y. Hoffman]{Ewa L. {\L}okas$^1$, Paul Bode$^2$
    and Yehuda Hoffman$^3$\\ $^1$Nicolaus Copernicus Astronomical Center, Bartycka 18,
    00-716 Warsaw, Poland\\ $^2$Princeton University Observatory, Princeton, NJ
    08544-1001, USA\\ $^3$Racah Institute of Physics, Hebrew University,
    Jerusalem 91904, Israel}

\begin{document}

\maketitle

\begin{abstract}
We use $N$-body simulations to measure mass functions in flat cosmological
models with quintessence characterized by constant $w$ with $w=-1$, $-2/3$
and $-1/2$. The results are compared to the predictions of the formula proposed
by Jenkins et al. at different redshifts,
in terms of FOF masses as well as Abell masses appropriate for direct comparison to
observations. The formula reproduces quite well
the mass functions of simulated haloes in models with quintessence. We use the
cluster mass function data at a number of redshifts from Carlberg et al.
to constrain $\Omega_0$, $\sigma_8$ and $w$. The best fit is obtained in the limit
$w \rightarrow 0$, but none of the values of $w$ in
the considered range $-1 \le w < 0$ can actually be excluded. However, the adopted value
of $w$ affects significantly the constraints in the $\Omega_0-\sigma_8$ plane.
Taking into account the dependence on $w$ we find $\Omega_0=0.32 \pm 0.15$ and
$\sigma_8=0.85_{-0.12}^{+0.38}$ ($68\%$ c.l.). Since
less negative $w$ push the confidence regions toward higher $\Omega_0$ and lower
$\sigma_8$ we conclude that relaxing the assumption of $w=-1$ typically made in such
comparisons may resolve the discrepancy between recent cluster mass function results
(yielding rather low $\Omega_0$ and high $\sigma_8$) and most other estimates.
The fact that high $w$ values are preferred may however also point towards some
unknown systematics in the data or the model with constant $w$ being inadequate.

\end{abstract}

\begin{keywords}
methods: $N$-body simulations -- methods: analytical -- cosmology: theory --
cosmology: dark matter -- galaxies: clusters: general -- large-scale structure of Universe
\end{keywords}

\section{Introduction}

Recently, our knowledge on background cosmology has improved
dramatically due to new supernovae and cosmic microwave background data.
Current observations favor a flat Universe with $\Omega_0\approx 0.3$
(see e.g. Harun-or-Rashid \& Roos 2001; Krauss 2003 and references therein) and the
remaining contribution in the form of cosmological constant or some
other form of dark energy. A class of models that satisfy these
observational constraints has been proposed by Caldwell, Dave, \&
Steinhardt (1998) where the cosmological constant is replaced with an
energy component characterized by the equation of state
$p/\varrho=w \neq -1$. The component can cluster on largest scales
and therefore affect the mass power spectrum (Ma et al. 1999)
and microwave background anisotropies (Doran et al.
2001; Balbi et al. 2001; Caldwell \& Doran 2003; Caldwell et al. 2003).

A considerable effort has gone into attempts to put constraints on models
with quintessence and presently the values of $-1 <w<-0.6$ seem most
feasible observationally (Wang et al. 2000; Huterer \& Turner 2001; Jimenez 2003).
Another direction of investigations is into the physical basis for the
existence of such component with the oldest attempts going back to Ratra
\& Peebles (1988). One of the promising models is based on so-called
``tracker fields" that display an attractor-like behaviour causing the
energy density of quintessence to follow the radiation density in the
radiation dominated era, but dominate over matter density after
matter-radiation equality (Zlatev, Wang, \& Steinhardt 1999; Steinhardt,
Wang, \& Zlatev 1999). It is still debated, however, how $w$ should depend
on time, and whether its redshift dependence can be reliably determined
observationally (Barger \& Marfatia 2001; Maor, Brustein, \& Steinhardt 2001;
Weller \& Albrecht 2001; Jimenez 2003; Majumdar \& Mohr 2003).

From the gravitational instability point of view, the quintessence field
and the cosmological constant play a very similar role: both can be
treated as (unclustered) dark energy components that differ by their
equation of state parameter, $w$.  Technically, the equations
governing the expansion of the Universe and the growth of density
perturbations in the two models differ only by the value of $w$.
Given the growing popularity of models with quintessence,
in this paper we generalize
the description of the mass functions to include the effect of dark
energy with constant $w$.

The mass function of clusters of galaxies has been used extensively
to estimate cosmological parameters, especially $\Omega_0$ and the
rms density fluctuation $\sigma_8$,
usually yielding (due to degeneracy)
a constraint on some combination of the two.
It has been shown that this
degeneracy can be significantly decreased by using the data at different
redshifts (Carlberg et al. 1997a; Bahcall, Fan \& Cen 1997; Bahcall \& Bode 2003)
due to different
growth rates of density fluctuations in different models. The growth rate
also varies among models with different $w$; one may therefore hope to
distinguish between them by analyzing mass function data at high redshift. It
is worth noting that the power spectra of models differing only by $w$
are different only at very large scales, so the rms density fluctuations at
scales of interest are almost identical when normalized to $\sigma_8$.

The paper is organized as follows.
In Section~2 we briefly summarize the properties of the cosmological
model with quintessence, including the linear growth factor of density
fluctuations. Section~3 presents the $N$-body simulations
and the mass function measured from their output. In Section~4 we describe the
transformation between mass functions in terms of FOF masses and in terms of Abell
masses.
Section~5 is devoted to the comparison of the theoretical mass functions to the data for
clusters at different redshifts to obtain constraints on cosmological parameters.
The Discussion follows in Section~6.

\section{The cosmological model}

Quintessence obeys the following equation of state relating its density
$\varrho_Q$ and pressure $p_Q$
\begin{equation}    \label{q1}
    p_Q = w \varrho_Q, \ \ \ \ {\rm where} \ \ -1 \le w < 0.
\end{equation}
The case of $w=-1$ corresponds to the usually defined cosmological
constant.

The evolution of the scale factor $a=R/R_0=1/(1+z)$ (normalized to unity
at present) in the quintessential Universe is governed by the
Friedmann equation
\begin{equation}       \label{th1}
    \frac{{\rm d} a}{{\rm d} t} = \frac{H_0}{f(a)}
\end{equation}
where
\begin{equation}    \label{th2}
    f(a) = \left[ 1+ \Omega_0 \left(\frac{1}{a}
    -1\right) + q_0 \left(\frac{1}{a^{1+3 w}} - 1\right) \right]^{-1/2}
\end{equation}
and $H_0$ is the present value of the Hubble parameter.
The quantities with subscript $0$ here and below denote the present
values. The parameter $\Omega$ is the standard measure of the amount of
matter in units of critical density and $q$
measures the density of
quintessence in the same units:
\begin{equation}    \label{q4}
    q  = \frac{\varrho_Q}{\varrho_{\rm crit}}.
\end{equation}
The Einstein equation for acceleration ${\rm d}^2 a/{\rm d} t^2 = -4 \pi G a (p
+\varrho/3)$ shows that $w < -1/3$ is needed for the accelerated
expansion to occur.

Solving the equation for the conservation of energy ${\rm d}
(\varrho_Q a^3)/{\rm d}a = -3 p_Q a^2$ with condition (\ref{q1}), we get the
following evolution of the density of quintessence in the general case
of $w=w(a)$:
\begin{equation}    \label{q2}
    \varrho_Q = \varrho_{Q,0} \exp \left[-3 \ln a + 3 \int_a^1 \frac{w(a)
    {\rm d} a}{a} \right]
\end{equation}
in agreement with Caldwell et al. (1998). For $w={\rm const}$, the case
considered in this paper, the formula reduces to
\begin{equation}    \label{q3}
    \varrho_Q = \varrho_{Q,0} \ a^{-3(1+w)}.
\end{equation}

The evolution of $\Omega$ and $q$ with scale factor (or equivalently redshift)
is given by
\begin{eqnarray}
    \Omega(a) &=& \frac{\Omega_0 f^2(a)}{a}, \label{th3} \\
    q(a) &=& \frac{q_0 f^2(a)}{a^{1+3 w}}  \label{th4}
\end{eqnarray}
while the Hubble parameter itself evolves so that $H(a) = H_0/[a \ f(a)]$.

\begin{figure}
\begin{center}
    \leavevmode
    \epsfxsize=8.5cm
    \epsfbox[40 40 340 300]{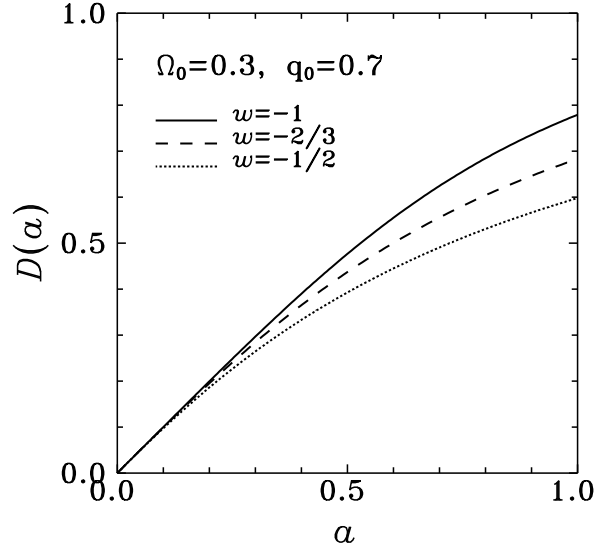}
\end{center}
    \caption{The linear growth rate of density fluctuations for
    $\Omega_0=0.3$, $q_0=0.7$ in three cases of $w=-1, -2/3$ and $-1/2$.}
\label{doda1}
\end{figure}

The linear evolution of the matter density contrast $\delta=\delta
\varrho/\varrho$ is governed by equation $\ddot{\delta} + 2 (\dot{a}/a) \dot{\delta}
- 4 \pi G \varrho \delta =0$, where dots represent derivatives with respect to time.
For flat models and arbitrary $w$ an analytical expression for $D(a)$ was found
by Silveira \& Waga (1994, corrected for typos). With our notation and the
normalization of $D(a)=a$ for $\Omega=1$ and $q=0$, it becomes
\begin{equation}    \label{th8}
    D(a) = a \ \ _2 F_1 \left[ -\frac{1}{3 w},
    \frac{w-1}{2 w}, 1-\frac{5}{6 w}, - a^{-3 w}
    \frac{1-\Omega_0}{\Omega_0} \right]
\end{equation}
where $_2 F_1 $ is a hypergeometric function. The solutions (\ref{th8}) for
$w=-1, -2/3$ and $-1/2$ are plotted
in Figure~\ref{doda1} for the cosmological parameters $\Omega_0=0.3$ and $q_0=0.7$.

\section{The simulated mass functions}

In order to study the cluster mass functions in models with quintessence we have used
a subset of the Gpc$^3$ dark matter $N$-body
simulations of Bode et al. (2001). Among the family of cosmological models they
considered two are of interest for us here: $\Lambda$CDM (with $w=-1$, $\Omega_0=0.3$,
$q_0=0.7$, $\Omega_{\rm b}=0.04$, $h=0.67$, $\sigma_8=0.9$) and QCDM
(with $w=-2/3$, $\sigma_8=0.84$, and all
other parameters the same). Both models had the primordial
spectral index $n=1$. The $w=-1$ and $w=-2/3$ simulations were carried out using the
Tree-Particle-Mesh (TPM) code described in Bode, Ostriker \& Xu (2000).

For the purpose of this study,
a new simulation with $w=-1/2$ (and $\sigma_8=0.88$,
with all the remaining parameters unchanged) was run using a
newer, publicly available version of TPM code which includes a number of
improvements over the earlier code (Bode \& Ostriker 2003). The changes in the
code which would affect the numerical results include improvements in the
time stepping (adding individual particle time steps within trees, and
a stricter time step criterion) and domain decomposition (adding an improved
treatment of tidal forces, and a new selection criterion for regions of
full force resolution). The changes in time step only affect the
innermost cores of collapsed objects, where the relaxation time is
short (Bode \& Ostriker 2003), leaving the mass function unchanged. The changes
in domain decomposition affect the mass function, but only at
the low-mass end:  in the $w=-1$ and $w=-2/3$ simulations, the mass function
is complete only above $6\times 10^{13} h^{-1}M_\odot$ (100 particles),
while in the $w=-1/2$ run it extends down to $10^{13} h^{-1}M_\odot$ (16
particles). In this paper we concentrate on the high-mass end of
the mass function, where these numerical differences have no impact.

The identification of halos in the final particle distribution has been performed
with the standard friends-of-friends (FOF) halo-finding algorithm with the linking
parameter $b=0.2$ (particles are linked if their separation is smaller than $b$ times
mean interparticle distance). As an alternative, we have also used
the HOP regrouping algorithm (Eisenstein \& Hut 1998). This procedure employs
several parameters; however, the resulting mass function is sensitive to only
one of them, the density (in units of the background density)
at the outer boundary of the halo, $\delta_{\rm outer}$.
As discussed by Eisenstein \& Hut (1998), assuming
$\delta_{\rm outer}=80$ is equivalent to using the linking parameter of $b=0.2$ in the
FOF halo-finding algorithm. Indeed, we verified that the mass functions measured in the
simulations are almost identical when using these two halo-finding schemes.

\begin{figure}
\begin{center}
    \leavevmode
    \epsfxsize=8.5cm
    \epsfbox[40 40 340 790]{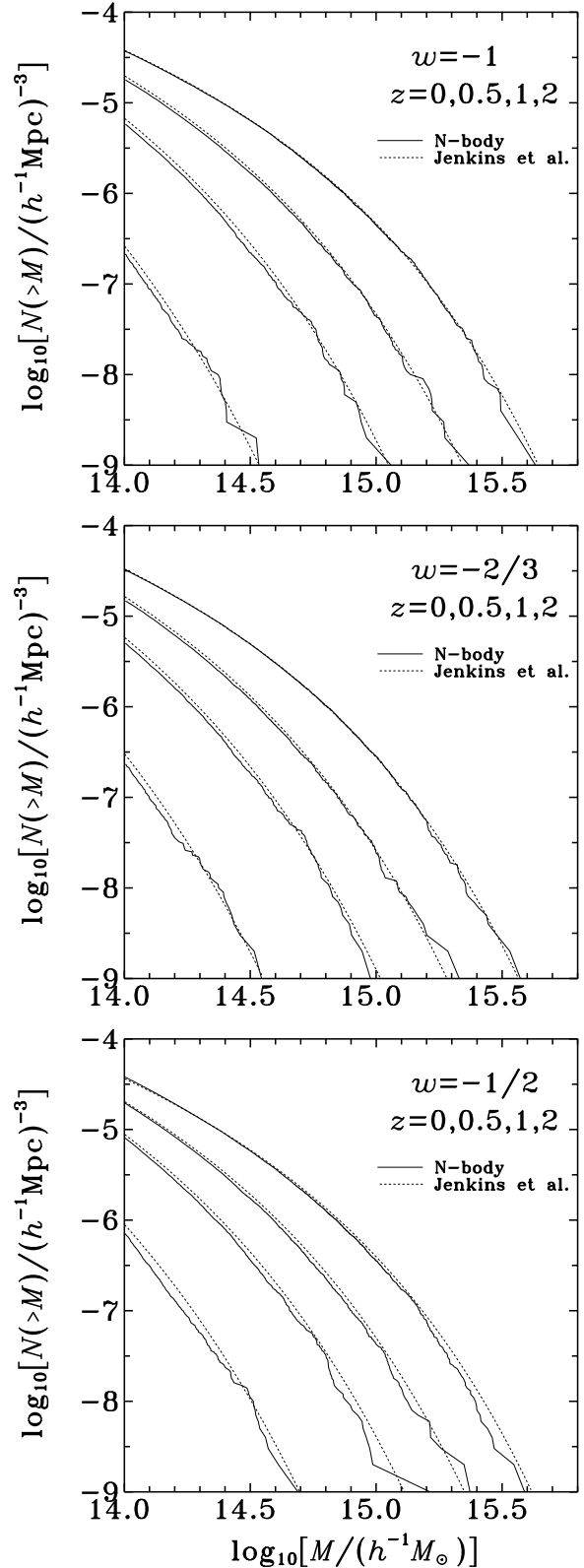}
\end{center}
    \caption{Comparison of the $N$-body cumulative mass functions (solid lines) obtained
    with $w=-1$ (upper panel), $w=-2/3$ (middle panel) and $w=-1/2$ (lower
    panel) to the predictions of the Jenkins formula
    (dotted lines) at different redshifts $z=0, 0.5, 1$ and $2$.}
\label{jesyz}
\end{figure}

Figure~\ref{jesyz} shows in solid lines the mass functions measured with FOF from $N$-body
simulations with $w=-1$ (upper panel), $w=-2/3$ (middle panel) and $w=-1/2$ (lower
panel) at redshifts $z=0, 0.5, 1$ and $2$.
The quantity plotted is the cumulative mass function (the comoving number density of
objects of mass grater than $M$) $N(>M) = \int_M^\infty n(M) {\rm d} M$,
where $n(M)$ is
the number density of objects with mass between $M$ and $M+{\rm d}M$.
Jenkins et al. (2001) established that when the masses are identified with FOF(0.2)
the simulated mass function in different cosmologies and at different epochs can be
well approximated by the following fitting formula
\begin{equation}   \label{q16}
   n_{\rm J}(M) = - \frac{\varrho_{\rm b}}{\sigma M}
   \frac{{\rm d} \sigma}{{\rm d} M} F(M)
\end{equation}
where
\begin{equation}   \label{q17}
   F(M) = 0.315 \exp(-|\ln \sigma^{-1} + 0.61|^{3.8}).
\end{equation}
In the above formulae $\varrho_{\rm b}$ is the background density, $\sigma$ is the
rms density fluctuation at top-hat smoothing scale $R$
\begin{equation}   \label{q13}
    \sigma^2 = \frac{D^2(a)}{(2 \pi)^3} \int {\rm d}^3 k P(k)
    W^2_{\rm TH}(k R)
\end{equation}
where $D(a)$ is the linear growth factor given by equation (\ref{th8}), $W_{\rm TH}(k R)$
is the top-hat filter in Fourier space and the mass
is related to the smoothing scale by $M=4 \pi \varrho_{\rm b} R^3/3$.

$P(k)$ in equation (\ref{q13}) is the power spectrum of density
fluctuations, which we assume here to be given in the form proposed by Ma
et al. (1999) for flat models. For the present time $(a=1)$ the power
spectrum is $P(k) = C k^n T^2 (k)$ where $n$ is the primordial power spectrum index
(we will assume $n=1$) and $T(k)$ is the transfer function. For the case of
cosmological constant ($\Lambda$CDM) we take the transfer function $T_\Lambda$
in the form proposed by Sugiyama (1995)
\begin{eqnarray}
   T^2_\Lambda(p) &=& \frac{\ln^2(1+2.34 p)}{(2.34 p)^2} \label{v4c} \\
   & \times & [1 + 3.89 p + (16.1 p)^2
   + (5.46 p)^3 + (6.71 p)^4]^{-1/2} \nonumber
\end{eqnarray}
where $p=k/(\Gamma h {\rm Mpc}^{-1})$ and $\Gamma=\Omega_0 h \exp [-\Omega_{\rm b}
(1+ \sqrt{2 h}/\Omega_0)]$.

For models with quintessence the transfer function is $T_Q = T_{Q\Lambda} T_\Lambda$,
where $T_{Q\Lambda}=T_Q/T_\Lambda$ can be approximated by fits given in Ma et al. (1999).
However, for $w \ne -1$, $T_{Q\Lambda}$ differs
from unity only at very large scales, i.e. very small wavenumbers.  Thus
if the spectra are normalized to $\sigma_8$ (the
rms density fluctuation smoothed with $R=8 h^{-1}$ Mpc), then
there is no need to introduce
the correction from $T_{\Lambda}$ to $T_Q$ in (\ref{q13}) because it does not affect
the calculation of $\sigma$.

The predictions for $N(>M)$ obtained from equations (\ref{q16})-(\ref{v4c}) for different $w$
and redshifts are shown in Figure~\ref{jesyz} as dotted lines. One can see that the Jenkins et al.
formula accurately reproduces the simulated mass function, especially at $z=0$. At higher
redshifts the formula of Jenkins et al. slightly overpredicts the number density of haloes;
a similar trend has been found recently by Reed et al. (2003). Therefore, we conclude that
although originally designed on basis of other cosmological
models, it can be considered valid also in the presence of quintessence. We therefore confirm
the result of Linder \& Jenkins (2003) who also found a good agreement between the simulations
and the predictions of Jenkins et al. (2001) formula in a different quintessence model.

\section{From FOF to virial and Abell masses}

The masses of clusters of galaxies are usually measured as the so-called Abell masses,
i.e. the masses inside the Abell radii of $1.5 h^{-1}$ Mpc. For the purpose of comparison with
observations we need to transform the mass functions expressed in terms of the
FOF masses to Abell masses. This can be done assuming a density distribution
inside the cluster. Since clusters are believed to be dominated by dark matter
(e.g. Carlberg et al. 1997b; {\L}okas \& Mamon 2003)
their density distribution can be well
approximated by the universal profile proposed by Navarro, Frenk \& White
(1997, NFW) for dark matter haloes
\begin{equation}    \label{c6}
     \frac{\rho(s)}{\rho_{\rm b} (z)} = \frac{\Delta_{\rm c} c^2 g(c)}{3 \Omega(z) s (1+ c s)^2}
\end{equation}
where the radius has been expressed in units of the virial radius $r_{v}$, $s=r/r_{\rm v}$.
The virial radius is defined as the distance from the
centre of the halo within which the mean density is $\Delta_{\rm c}$ times the
critical density, $\rho_{\rm crit}$. The value of the virial
overdensity $\Delta_{\rm c}$ is estimated from the spherical collapse model and depends
on the cosmological model. For flat models with constant quintessence parameter $w$, its
behaviour can be well approximated as (Weinberg \& Kamionkowski 2003)
\begin{equation}    \label{deltac}
	\Delta_{\rm c} = 18 \pi^2 (1 + a \, \theta^b) \Omega(z)
\end{equation}
with $\theta = (1/\Omega_0 - 1) (1 + z)^{3 w}$ and
$a = 0.399 - 1.309 [(-w)^{0.426} - 1]$, $b = 0.941 - 0.205 [(-w)^{0.938} - 1]$.

\begin{figure}
\begin{center}
    \leavevmode
    \epsfxsize=8.5cm
    \epsfbox[40 40 340 300]{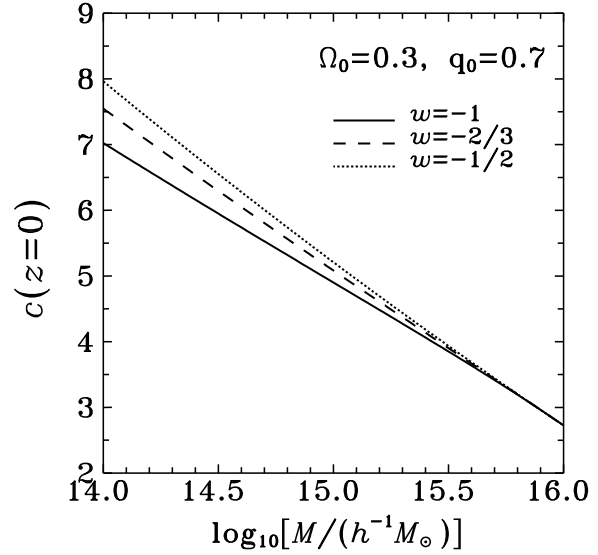}
\end{center}
    \caption{The concentration parameter at $z=0$ calculated from the model of
    Bullock et al. (2001) for $\Omega_0=0.3$, $q_0=0.7$ and $w=-1, -2/3$ and $-1/2$ with
    the remaining parameters as in the simulations, except that we kept $\sigma_8=0.9$
    for all cases.}
\label{cm}
\end{figure}

The quantity $c$ introduced in equation (\ref{c6})
is the concentration parameter, $c=r_{\rm v}/r_{\rm s}$, where $r_{\rm s}$ is the scale radius
(at which the slope of the profile is $r^{-2}$). The function $g(c)$ in equation
(\ref{c6}) is $g(c) = 1/[\ln (1+c) - c/(1+c)]$. From cosmological $N$-body simulations
(NFW; Jing 2000; Jing \& Suto 2000; Bullock et al. 2001), we know that $c$ depends on the
mass and redshift of formation of the object, as well as the initial power spectrum of density
fluctuations. We will approximate this dependence using the toy model of Bullock et al. (2001)
(their equations (9)-(13) with parameters $F=0.001$ and $K=3.0$ as advertised for masses
$M > 10^{14} h^{-1} M_{\sun}$). The predictions of the model at $z=0$ for
cosmological models as in our simulations (but all normalized to
$\sigma_8=0.9$) are shown in Figure~\ref{cm} for masses $10^{14}-10^{16} h^{-1} M_{\sun}$.
For higher redshifts we assume
$c(z) = c(z=0)/(1+z)$, following Bullock et al. (2001). The dependence of the concentration
on $w$ is rather weak but there is a trend of larger $c$ values for less negative $w$.
A similar trend was observed in the properties of dark haloes obtained in the
$N$-body simulations by Klypin et al. (2003), although for smaller masses.

Since the NFW profile describes the halo properties in terms of the virial radius and
the corresponding virial mass $M_{\rm v} = 4 \pi r_{\rm v}^3 \Delta_{\rm c} \rho_{\rm crit}/3$, we have to
transform the FOF masses first to virial masses. As previously noted, the FOF parameter $b=0.2$
corresponds to the local overdensity at the border of the halo $\delta_{\rm outer} = 80$.
Therefore we will assume that the FOF masses are equal to masses enclosed by such isodensity
contour, $M_{\rm FOF} = M_{80}$. Using the NFW distribution we find that for our simulated
$\Lambda$CDM model at $z=0$ we have $M_{80} \approx M_{\rm v}$, while for higher redshifts
$M_{80} < M_{\rm v}$. In the case of models with $w=-2/3$ we instead get $M_{80} > M_{\rm v}$, and more
so for $w=-1/2$. The differences between $M_{80}$ and $M_{\rm v}$ for the models, redshifts
and mass range considered here are typically of the order of a few percent and
do not exceed 20\%. They depend somewhat on mass, and are due mainly to the
dependence of $\Delta_{\rm c}$ on cosmology.
Once the FOF masses are translated to virial masses we can obtain the Abell masses $M_{\rm A}$
using the NFW mass distribution following from equation (\ref{c6})
\begin{equation}    \label{c8}
    M_{\rm A} = M(s_{\rm A}) = M_{\rm v} g(c) \left[ \ln (1+c s_{\rm A}) - \frac{c s_{\rm A}}{1 + c
    s_{\rm A}}  \right]
\end{equation}
where $s_{\rm A}=r_{\rm A}/r_{\rm v}$ is the Abell radius in units of the virial radius.

\begin{figure}
\begin{center}
    \leavevmode
    \epsfxsize=8.5cm
    \epsfbox[40 40 340 790]{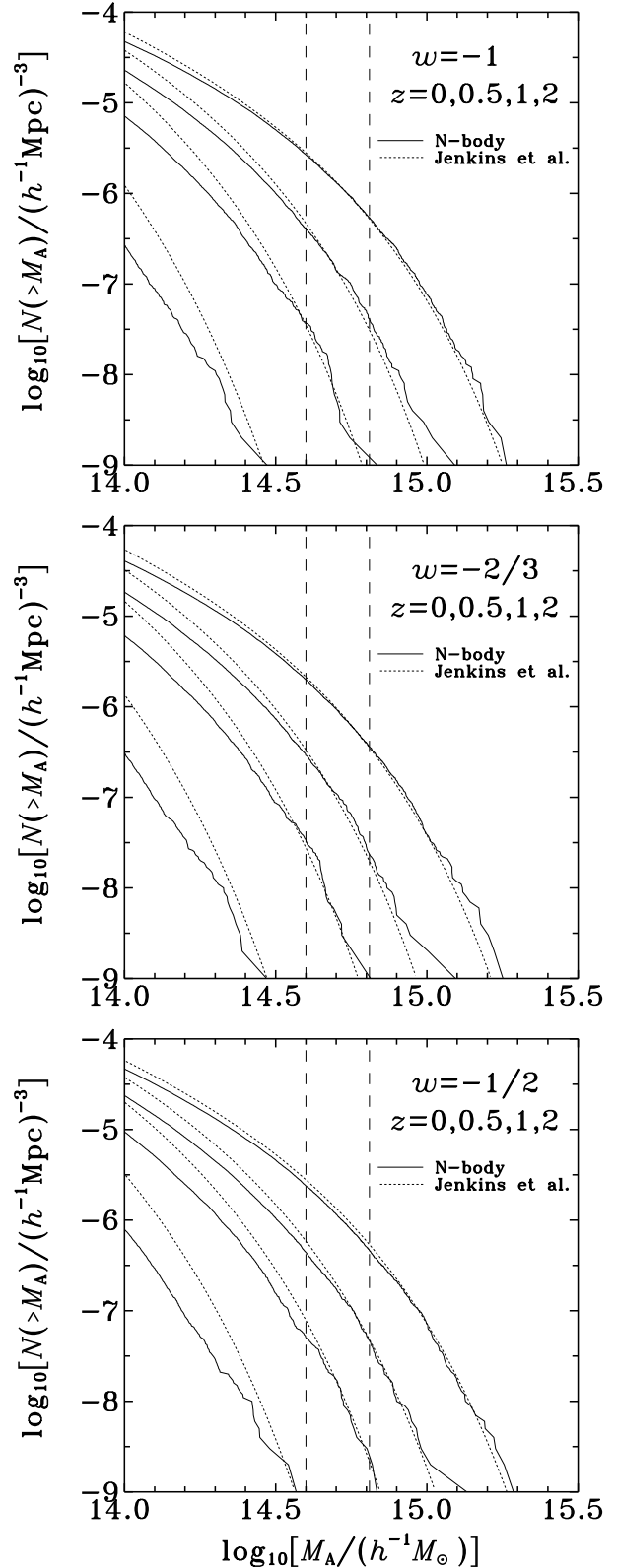}
\end{center}
    \caption{Cumulative mass functions in terms of Abell masses measured in $N$-body
    simulations (solid lines) and obtained from the predictions of the Jenkins formula
    with mass transformation (dotted lines) for
    $w=-1$, $w=-2/3$ and $w=-1/2$ and at different redshifts. The two vertical dashed lines
    in each panel indicate the mass range of the data used in Section~5.}
\label{asymza}
\end{figure}

Figure~\ref{asymza} shows the cumulative mass functions in terms of the Abell masses as
measured in the simulations (solid lines) and as calculated with the Jenkins et al. formula
(\ref{q16})-(\ref{q17}) with the proper mass transformation (dotted lines). The agreement is
very good for $z=0$.
In general, the analytic predictions match the $N$-body results as long
as the virial radius is near or greater than the Abell radius.
However, at
higher redshifts the predictions for lower masses are overestimated.
Several effects might lead to this disagreement.  Abell masses were
measured in the simulations in the manner described in Bode et al. (2001).
In this method,
clusters cannot be closer together than 1 $h^{-1}$Mpc; also, if two
Abell radii overlap, particles in the overlap region are only included
in one of the clusters (based on binding energy).  These two factors
may lead to fewer objects -- lower mass objects being subsumed into
larger ones.

As discussed by Bode et al. (2001) higher resolution simulations produce somewhat
steeper Abell mass functions which would agree better with the predictions. We must also keep in
mind that the result was obtained with the assumption of NFW profile, while for smaller
haloes this may not be the case. The haloes of mass $10^{14} h^{-1} M_{\sun}$ have only 160
particles in our simulations and can hardly be expected to have a well defined density profile.
We have also extrapolated the model for concentration of Bullock et al. (2001) both for
quintessence and very large masses, two regimes where it has never been tested, while even
in well studied models concentration shows substantial scatter. Given the number
of approximations involved in the result we conclude that the fits are satisfactory, especially
in the mass range between the two vertical dashed lines in each panel of Figure~\ref{asymza},
covered by the data, which will be used in the next Section.

\section{Comparison with observations}

\begin{figure}
\begin{center}
    \leavevmode
    \epsfxsize=8.5cm
    \epsfbox[40 40 340 300]{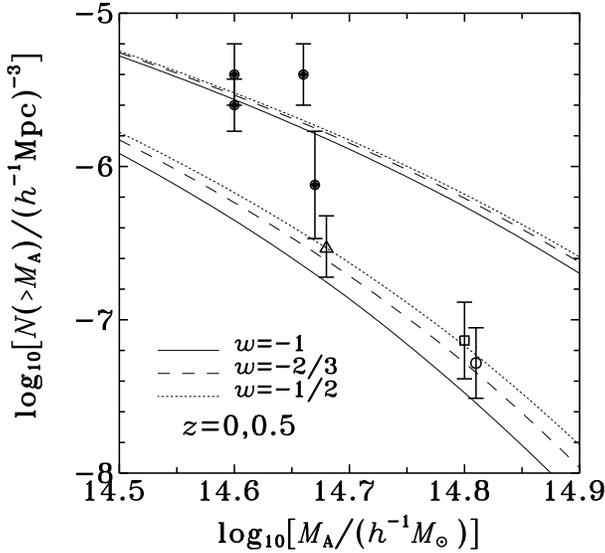}
\end{center}
    \caption{The data of Carlberg et al. (1997a) for $\Omega_0=0.3$
    versus Abell mass functions predicted for different
    models with quintessence. The filled circles mark the four data points at low redshift
    $0 < z <0.1$. The open symbols correspond to higher redshift data:
    $0.18 < z <0.35$ (open triangle), $0.35 < z <0.55$ (open square)
    and $0.55 < z <0.85$ (open circle). The lines show the models i.e. our
    mass functions in terms of Abell mass discussed in the previous Section. All
    models have $\Omega_0=0.3$, $q_0=0.7$ and $\sigma_8=0.9$
    but different $w$: $w=-1$ (solid lines),
    $w= -2/3$ (dashed lines) and $w=-1/2$ (dotted lines). The three upper lines are for $z=0$,
    while the three lower ones for $z=0.5$.}
\label{data}
\end{figure}

For comparison of the theoretical mass functions with observations we used the data
for cluster mass function from Carlberg et al. (1997a). The data consist of seven data
points of $N(>M)$ for different Abell masses $M > 4 \times 10^{14} h^{-1} M_{\sun}$,
for redshift bins in the range $0 < z < 0.85$. There are four data points coming
from different surveys in the low
redshift bin $0 < z <0.1$ and  three data points in higher redshift
bins $0.18 < z <0.35$, $0.35 < z <0.55$
and $0.55 < z <0.85$, respectively (see Table~1 of
Carlberg et al. 1997a). The data for $\Omega_0=0.3$ are shown in Figure~\ref{data}
with the filled circles corresponding to the low-redshift samples and open symbols to
the higher-redshift ones.

Together with the data we show in Figure~\ref{data}
our model Abell mass functions discussed in the previous
Section. All models have $\Omega_0=0.3$, $q_0=0.7$, $\sigma_8=0.9$ and
differ only in the value of $w$. The three upper lines shown in the Figure are for $z=0$,
while the three lower ones for $z=0.5$. We note that the small difference between the
models with different $w$ at $z=0$ is due only to our mapping procedure between the FOF and
Abell masses which involves $w$-dependent parameters $\Delta_{\rm c}$ and $c$. (The predictions
of the Jenkins et al. 2001 formula for FOF masses applied without any corrections
would be identical for the given set of cosmological parameters since the differences in
the power spectrum for different $w$ are negligible.) The differences between predicted
Abell mass functions for different $w$ start to be more pronounced at higher redshifts
(three lower curves in Figure~\ref{data}) due to different growth factors of density
perturbations in these models (see Figure~\ref{doda1}).

We performed a standard
$\chi^2$ fit to the data points in terms of $\log N(>M)$ with three free parameters:
$\Omega_0$, $\sigma_8$ and $w$. Since the data are in the form of a
$\log N(>M)$ value per
redshift bin, for the predictions we take the mean redshift of
the bin (it is not necessary to average over redshift because $\log N(>M)$ is
approximately linear in redshift). When considering different $\Omega_0$ in
models we adjust the data point by linearly interpolating between the data
given in Carlberg et al. (1997a) for $\Omega_0=0.2$ and $\Omega_0=1$.
The analysis has been done only for flat models, i.e. we kept
$q_0=1-\Omega_0$. We also assumed Hubble constant $h=0.7$ and the primordial
spectral index $n=1$.

\begin{figure*}
\begin{center}
    \leavevmode
    \setbox100=\hbox{\epsfxsize=18cm\epsffile{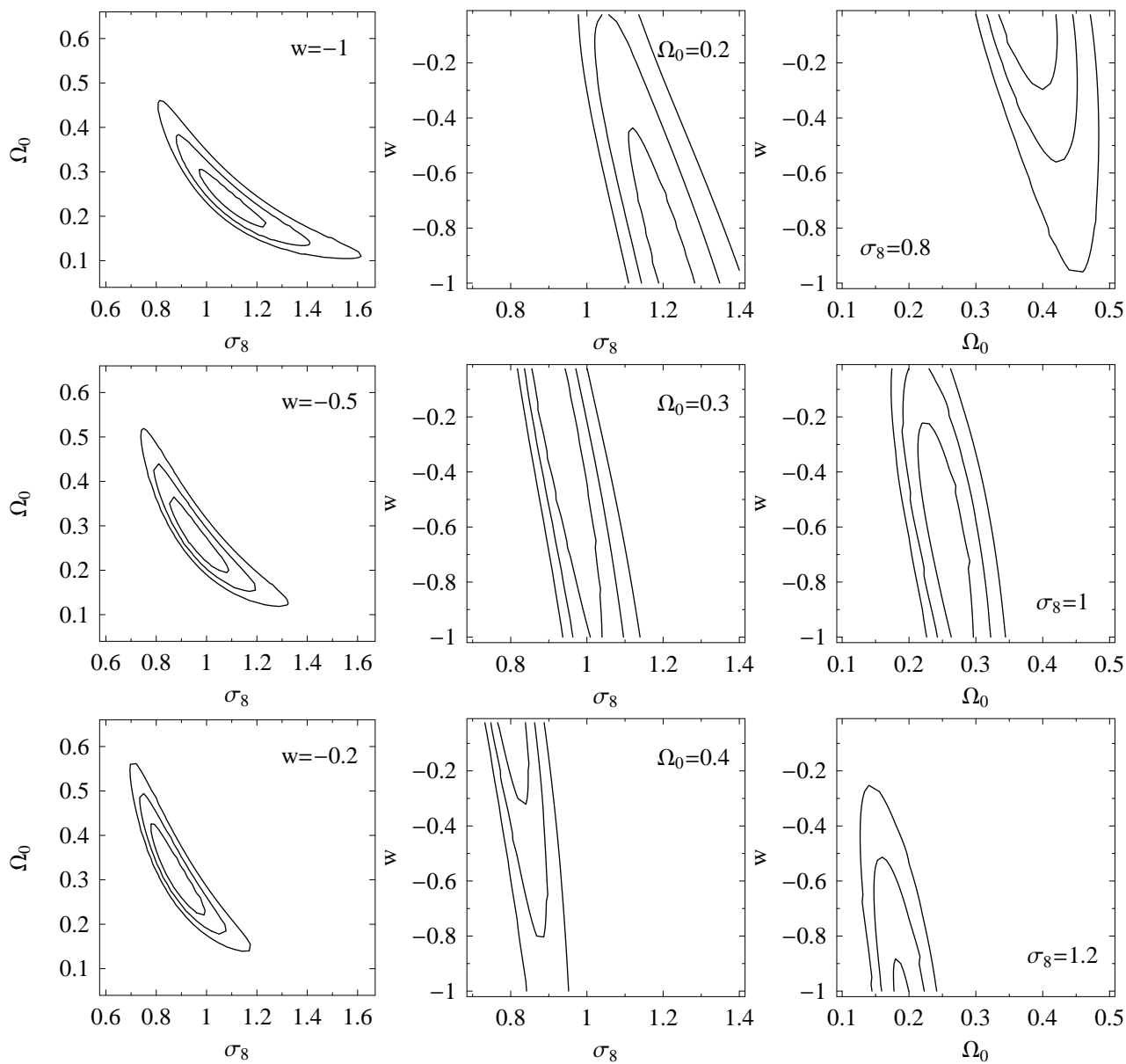}}
    \hbox{\box100}
\end{center}
    \caption{The $1\sigma$, $2\sigma$ and $3\sigma$ probability contours in the
    $\sigma_8-\Omega_0$ (left column),  $\sigma_8-w$ (middle column) and
    $\Omega_0-w$ plane (right column) resulting from the fits to the cluster mass
    function data of Carlberg et al. (1997a). The assumed value of the third parameter is given
    in the corner of each panel.}
\label{contours}
\end{figure*}

Figure~\ref{contours} shows the $1\sigma$, $2\sigma$ and $3\sigma$
probability contours in the $\sigma_8-\Omega_0$ (left column), $\sigma_8-w$ (middle column)
and $\Omega_0-w$ (right column) parameter planes respectively. For each plane the cuts
through the confidence region are done for three values of the third parameter;
the value is indicated
at the corner of each panel. The contours correspond to $\Delta \chi^2 = \chi^2 -
\chi^2_{\rm min} = 3.53, 8.02, 14.2$, where the minimum value $\chi^2_{\rm min}
=10.1$ is obtained for $\Omega_0=0.32$, $\sigma_8=0.85$ and $w \rightarrow 0$.

The contours in the  $\sigma_8-\Omega_0$ plane presented in the left column of Figure~\ref{contours}
are the cuts through the confidence region at (from top to bottom) $w=-1$, $w=-0.5$ and $w=-0.2$.
They show a typical shape obtained in this kind of analyses.  However, the three cuts through the
confidence space shown in this column of Figure~\ref{contours}
actually move significantly when the assumed value of $w$ is changed (note
that the axes scales in the left column of the Figure are the same in each panel).
Taking into account the dependence on $w$ and the variability of the contours in the whole
range $-1 \le w < 0$, we find $\Omega_0=0.32 \pm 0.15$ and
$\sigma_8=0.85_{-0.12}^{+0.38}$  at $68\%$ confidence level. While for
$w=-0.2$ the best-fitting values of the remaining parameters are $\Omega_0=0.3$, $\sigma_8=0.9$,
for lower $w$ the contours move towards lower $\Omega_0$ and higher $\sigma_8$. For $w=-1$
they are centered on $\Omega_0 \approx 0.2$, $\sigma_8 \approx 1.1$.

The middle and right columns of Figure~\ref{contours} show the constraints on $w$ in two planes,
$\sigma_8-w$ and $\Omega_0-w$ respectively. The middle column has the cuts for $\Omega_0=0.2$,
$\Omega_0=0.3$ and $\Omega_0=0.4$ from the top to the bottom panel, while in the right column
the values of the third parameter are $\sigma_8=0.8, 1$ and $1.2$.
As can be seen in the plots, there is a strong
degeneracy between $w$ and any of the two remaining parameters. The particular shape of the
contours can be understood by referring back to Figure~\ref{doda1}, showing the growth rate
of density fluctuations for different $w$. By changing the normalization of the curves
in Figure~\ref{doda1} to give the same value of $D(a)$ at present ($a=1$),
it is easily seen that the magnitude of density fluctuations
drops faster with redshift for more negative $w$. Thus in order to reproduce a given
redshift dependence of the cluster mass function data, for a more negative $w$ and a
given $\Omega_0$ ($\sigma_8$) a higher value of
$\sigma_8$ ($\Omega_0$) is needed (as both these parameters enhance the growth of structure).

\section{Discussion}

The confidence regions we obtained in Figure~\ref{contours} in the $\sigma_8-\Omega_0$ plane
are qualitatively similar to the
results of the analysis of the same data by Carlberg et al. (1997a),
which differed from
our approach in that they used the Press \& Schechter (1974) approximation, a power-law
distribution of mass in clusters, and did not consider the dependence on $w$.
While for $w=-0.2$ the best-fitting values of the remaining parameters
are $\Omega_0=0.3$, $\sigma_8=0.9$
in very good agreement with other estimates (e.g. from CMB, see Spergel et al. 2003 or type Ia
supernovae, see Tonry et al. 2003), for
lower $w$ the contours move towards lower $\Omega_0$ and higher $\sigma_8$.
At $w=-1$ they are centered on $\Omega_0 \approx 0.2$, $\sigma_8 \approx 1.1$.

A very similar methodology is provided by using  Sunyaev-Zeldovich cluster
surveys to probe the evolution of the surface density of clusters as a
function of redshift (Battye \& Weller 2003 and references therein). The constraints
derived by the Sunyaev-Zeldovich surveys are based on the same physical processes
and models as those employed here, namely the mass spectrum of haloes in the
quintessential cosmology.
Battye \& Weller  (2003) studied the constraining power of  Sunyaev-Zeldovich surveys
by analyzing mock catalogs of future surveys. Comparing the results obtained here and
by Battye \& Weller we find that apart from the future PLANCK survey
the Carlberg et al. (1997a) data provide tighter or comparable constraints to
the Sunyaev-Zeldovich surveys. The PLANCK cluster survey will reduce the
uncertainties in $\sigma_8$ and $\Omega_0$ by roughly a factor of 2 in comparison with
the present results.

Assuming the $\Lambda$CDM cosmological model,
Bahcall et al. (2003) used data from the SDSS collaboration
to estimate that for $\Omega_0 \approx 0.2$, $\sigma_8 \approx 0.9$,
a lower value than found here.  However, the current result includes
higher redshift data, which leads to higher estimates for $\sigma_8$,
independent of $\Omega_0$ (Bahcall \& Bode 2003).
It seems that assuming $w=-1$ when considering cluster abundances
leads to rather low values of $\Omega_0$ (or alternatively,
high values of $\sigma_8$);  as
pointed out by Oguri et al. (2003), these are in mild conflict
with most other estimates.
Oguri et al. (2003)
suggested that decaying cold dark matter may resolve this discrepancy. Our results offer another
possibility: they show that the
constraints on $\Omega_0$ and $\sigma_8$ from cluster mass functions are in better agreement
with other estimates if the assumption of $w=-1$ is relaxed and a less negative value of
$w$ is adopted.

We have demonstrated the existence of a strong
degeneracy between $w$ and any of the two remaining parameters, $\Omega_0$ and $\sigma_8$,
which is not broken in spite of using high redshift data: reproducing the evolution of the
cluster mass function data requires a higher value of $\Omega_0$ or $\sigma_8$ for more negative $w$.
A similar behaviour of the confidence regions, including the degeneracies, in the
$\sigma_8-w$ and $\Omega_0-w$ planes
was recently observed also by Schuecker et al. (2003, see the lower panels of their Fig. 3)
who used the REFLEX X-ray cluster sample.
As discussed by Douspis et al. (2003) and Crooks et al. (2003) the degeneracies between $w$
and other cosmological parameters also plague the constraints obtained from other, e.g.
CMB, data sets.

The best fit to the cluster mass function data is obtained for a surprisingly
high value of $w \approx 0$, but
the dependence on $w$ is rather weak and no value in the entire considered range
$-1 \le w < 0$ can actually be excluded: for every value in this range
a reasonable combination of $\Omega_0$ and $\sigma_8$ can be found which places the point in
$1\sigma$ confidence region. It is interesting to note, however, that depending on the values of
$\Omega_0$ and $\sigma_8$, we get upper or lower limit on $w$: for high $\Omega_0$
and low $\sigma_8$ the contours tend to provide a lower limit on $w$, while for low $\Omega_0$
and high $\sigma_8$ we have an upper limit. Combined with additional constraints on $\Omega_0$ or
$\sigma_8$ from other data sets, cluster mass functions can therefore prove useful
in estimating the value of $w$. Such an analysis has been recently performed by Schuecker et al.
(2003) who combined the data from X-ray clusters with the data for SNIa. Since the supernova
data show a strong preference for negative $w$ the resulting best-fitting value of $w$ is very
close to $-1$.

With the estimates of the cosmological parameters presently available
the analysis presented here tends to provide a lower limit on $w$. For example, for the best
estimates from WMAP (Spergel et al. 2003), $\Omega_0=0.27$ and $\sigma_8=0.9$, the constraint from
our analysis is $-0.5 < w < 0$ at 95\% confidence level. The values of $\Omega_0$ and
$\sigma_8$, however, are not yet known exactly so a proper combination with other data sets would
have to be performed.
Typically, other data sets give an upper limit on $w$, e.g. WMAP in combination with other
astronomical data gives $w < -0.78$ (Spergel et al. 2003) while from the SNIa data Tonry et al. (2003)
obtain $w < -0.73$. This means that the combination of these limits with the cluster mass function
data will probably result in a preferred range of
$w$ instead of only an upper or lower limit. The high values of $w$ preferred by the cluster
mass function data may also turn out to be in conflict with other estimates. This
would point towards some unknown systematics in the data or inconsistencies in the models.
Then the cosmological model with constant $w$ would have to be rejected and replaced with
one involving some time-dependence of $w$.

\section*{Acknowledgements}

We thank Ofer Lahav, Gary Mamon and the anonymous referee for their
comments on the paper.
This work was supported in part by the Polish KBN grant 2P03D02726.
Computer time to perform $N$-body simulations was provided by NCSA.
PB received support from NCSA NSF Cooperative Agreement ASC97-40300,
PACI Subaward 766. YH acknowledges support from
the Israel Science Foundation (143/02) and the Sheinborn Foundation.

\end{document}